\documentstyle[epsf,epsfig,aps]{revtex}
\renewcommand{\thefootnote}{\fnsymbol{footnote}}
\begin{document}
\newcommand{\be}{\begin{eqnarray}}
\newcommand{\dlq}{\lq\lq}
\newcommand{\ee}{\end{eqnarray}}
\newcommand{\ben}{\begin{eqnarray*}}
\newcommand{\een}{\end{eqnarray*}}
\newcommand{\beq}{\begin{equation}}
\newcommand{\eeq}{\end{equation}}
\renewcommand{\baselinestretch}{1.0}
\newcommand{\as}{\alpha_s}
\def\eq#1{{Eq.~(\ref{#1})}}
\begin{flushright}
TAUP   2650-2000\\
\today \\
\end{flushright}
\vspace*{1cm}
\setcounter{footnote}{1}
\begin{center}
{\Large\bf  Soft Pomeron in QCD}
\\[1cm]
 Eugene
Levin  \\ ~~ \\

{\it $^3$ HEP Department, School of Physics and Astronomy } \\
{\it Tel Aviv University, Tel Aviv 69978, Israel } \\ 
{\it and}\\
{\it DESY Theory,  22603 Hamburg, Germany}
~~ \\ ~~ \\
\centerline{\it Talk, given at Diffraction'2000, Sept. 2 - 7, Centraro,
Italy}

~~\\
~~\\

\end{center}
\begin{abstract}
This talk is a brief presentation of   our view on the Pomeron, as a
non-perturbative QCD phenomenon  but from sufficiently short distances.
Our approach is based on the scale anomaly of QCD and emphasizes the
r{\^o}le of semi--classical QCD vacuum fields. We show that both the
intercept and the slope of Pomeron trajectory appear to be determined by
the energy density of non-perturbative QCD vacuum. The particular example
of  semi--classical QCD vacuum fields is discussed based on a  new type of
instanton--induced
interactions (``instanton ladder'') that leads to the rising with  
energy cross section $\sigma \sim s^{\Delta_P}$ of Regge type (the   
Pomeron).

\end{abstract}

\renewcommand{\thefootnote}{\arabic{footnote}}
\setcounter{footnote}{0}

\section{Main idea}

It is well known that the Pomeron structure is  one of the most
challenging problem of QCD. We  need soft Pomeron to describe
the experimental data on soft processes\cite{DL} but we have no idea
why  a  Regge pole with the intercept close to unity, which we
call soft Pomeron, could appear in our microscopic theory-QCD. Our main
idea \cite{SAP} is that 
{\it soft Pomeron\,\,\, 
$\longrightarrow$\,\,\,  non-perturbative QCD but at sufficiently
short
distances.
  }

Indeed,  pQCD calculations\cite{BFKL}  lead to so called BFKL Pomeron
which is not a Regge pole. On the other hand, perturbative approach for
1+2 dimensional QCD \cite{BFKL2} gives a reggeon with the intercept larger
than unity if we introduce a scale in our theory ( gluon mass ). For
massless 1+2 QCD we still do not have a Pomeron. The lesson which we
learned from this approach is that we have to find 
the origin of a violation of the scale invariance in QCD which leads to an
appearance of a typical scale in massless QCD.  This momentum scale should
be large enough since our phenomenological Pomeron has  sufficiently large
typical momentum. Indeed, (i) the slope of the Pomeron trajectory
 $ \alpha_P (t) = 1 + \Delta_P +\alpha'_P (0)
\,t$ \, is equal to 
 $ \alpha'_P (0)\,=\,0.25 \,GeV^{-2} \,\ll\,\,\alpha'_R
 (0)\,=\,1 \,GeV^{-2}$, where $\alpha'_R$ is the slope of secondary
trajectories; and (ii)
the $t$ - slope of the triple Pomeron vertex $G_{3P} (t)$ is
very
small.

Therefore, we have three principle questions to answer:(i) why we have a
Pomeron in QCD; (ii)  why the Pomeron intercept is so small ($\Delta_P
\approx 0.08 \div 0.1 $ ) for
non-perturbative QCD and (iii)  why  a typical momentum scale is so high ($
M_0 \,\propto\,1/\alpha'_P \,\approx\, 
\, 2\,GeV$ )  for 
 ``soft" Pomeron.

\section{ High momentum scale for Pomeron}
The first question which we would like to answer is
 { \it does sufficiently
large
mass  scale $M_0 \,\,\approx\,\, \,\,\,\,\,2
\,GeV $  appear in
non-perturbative QCD?}. Our answer is {\it yes}.

As have been mentioned,  the key ingredient of our approach is the
breakdown
of the scale invariance in QCD, reflected in scale anomaly \cite{SA}.
QCD is scale invariant on the classical tree level in the chiral limit of
the massless quarks. However, this invariance is broken due to scale 
dependence
of the QCD coupling constant, which introduces a
dimensionful scale ( $\Lambda$ ). On the formal level, the breakdown of
scale invariance in the theory is reflected by non-conservation of scale
current, and thus in the non-zero trace of the  the energy - momentum
tensor $\theta^{\mu}_{\mu}$ \cite{SA1}. Scale anomaly leads to a set of
powerful low energy theorems\cite{NSVZ}, from which we are going to use 
\beq \label{T1}
i \int dx\, \langle\, T\, \left\{ \Theta(x)\,\Theta(0) \right\}
\rangle\,   
= \,\int \frac{dM^2}{M^2}
[\rho_\Theta^{\rm phys}(M^2)-\rho_\Theta ^{\rm pt}(M^2)]
\,= \,- 4  \langle 0 | \Theta | 0  \rangle
\,= \,- 16 \epsilon_{vac} \neq 0 
\eeq

Let us make a simple estimate counting the number of coupling constant
in \eq{T1}. On the l.h.s. we have each term of the order of $\alpha^2_S$
while on the r.h.s. $\epsilon_{vac} \propto \alpha_S$. Therefore, \eq{T1}
holds only because the range of integration over $M^2$ is of the order of
$1/\alpha_S$ or, in other words, typical $M^2 \propto 1/\alpha_S$. This is
a reason for large momentum scale in QCD. The real estimates for the value
of $M^2_0$ has been performed using the chiral limit of QCD in
which\cite{CL}
\beq \label{CHIRAL}
\Theta_\mu^\mu\, =\,
-\partial_\mu \pi^a \partial^\mu\pi^a +2 m_\pi^2 \pi^a \pi^a + \cdots 
\,\,\,\,\,\,\mbox{and}\,\,\,\,\,\,
\langle{\pi^+\pi^-} |  \theta_\mu^\mu | 0 \rangle  = M^2 = ( \alpha_S)^0
\,\,.
\eeq

Substituting \eq{CHIRAL} in \eq{T1} one can get the estimates for the
value of typical mass, namely
\beq \label{M0}
M_0^2 \, \, \simeq \, \, 32 \pi \left\{ \frac{|\epsilon_{\rm
vac}|}{  N_f^2 - 1}\right\}^{\frac{1}{2}}
 \,\,=\,\,4\,\div\,6 \,\,GeV^2 \,\approx\,\,1/\alpha'_P
\,\ll\,\,\Lambda \,\,,
\eeq
for  $\epsilon_{vac} \approx - (0.24\,\,GeV )^4$ .
The physical way of understanding the disappearance of the coupling
constant in $\Theta_\mu^\mu \propto g^2 F^2$ is to assume that the
non-perturbative QCD vacuum  is  dominated by the semi-classical
fluctuations of the gluonic fields with $F \propto 1/g$.

\section{Scale anomaly of QCD $\mathbf{ \longrightarrow}$
Soft Pomeron} 
Armed with this knowledge on QCD vacuum we can formulate our strategy
for searching of non-perturbative Pomeron contribution. 

\begin{figure}[htbp]
\begin{center}
\epsfig{file=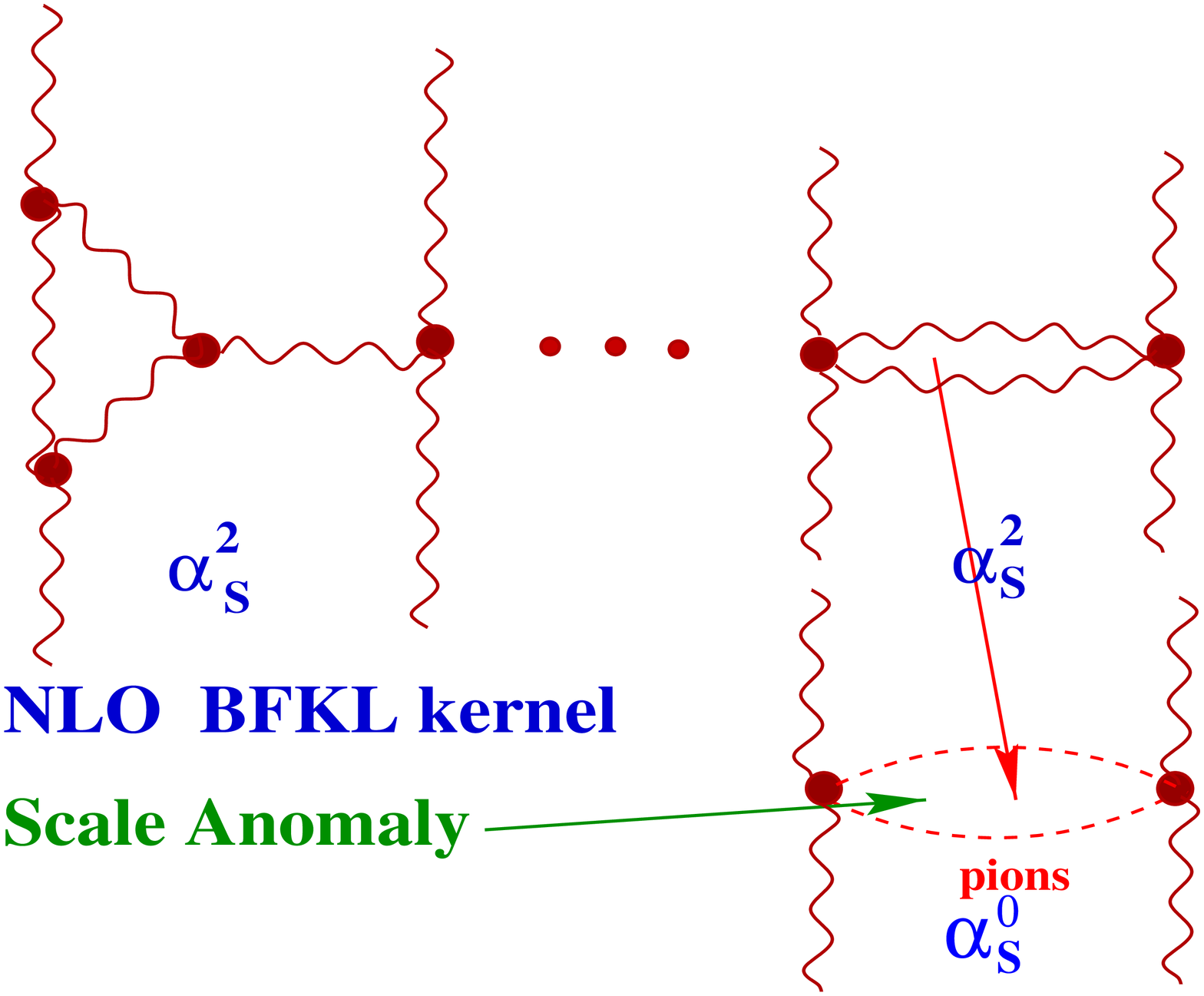,width=12cm,height=6cm}
\end{center}
\caption{}
\end{figure}
 It consists of
two steps: (i) first is to find the  $(\Theta_\mu^\mu)^2 $
 contribution to the next-to-leading order BFKL kernel
\cite{NLOBFKL} which is formally  $\sim\,O(\alpha^2_S)$; (ii) second to
replace this contribution by the non-perturbative one which is of the
order of $\alpha^0_S $ due to scale anomaly. These steps are shown in
Fig.1

In doing so, we obtained that the total cross section is proportional to
$s^{\Delta_P}$ with
\beq \label{DELTA}
\Delta_P \,\,=\,\frac{\pi^2}{2} \,
\left( \frac{8 \pi}{b}
\right)^2\,\frac{18}{32 \pi^2}\,\int\,\frac{ d M^2}{M^6}
\,\left(\,\rho^{phys}_{\theta}( M^2 )\,\,-\,\,\rho^{pQCD}_{\theta}( M^2
)\,\right) \,=\, \frac{1}{48} \,\ln\frac{M^2_0}{4 m^2_{\pi}}\,\,;
\eeq
where we use \eq{CHIRAL} for the non-perturbative contribution to
$\Theta_\mu^\mu$. Numerically, \eq{DELTA} gives $ \Delta_P = 0.08 \div
0.1$\cite{SAP} for $M^2_0 = 4 \div 6 \,GeV^2$  
in a good agreement with the experimental data \cite{DL}.

Therefore, we obtained a soft Pomeron that we needed. All corrections to
our approach are small since they are proportional to $\alpha_S(M^2_0) \ll
1$. However, we would like to stress that we cannot control all other
contributions which stem from long distances $r \gg  1/M_0$.  We believe
that they are irrelevant to the Pomeron structure.

\begin{figure}[htbp]
\begin{tabular}{c c}
\epsfig{file=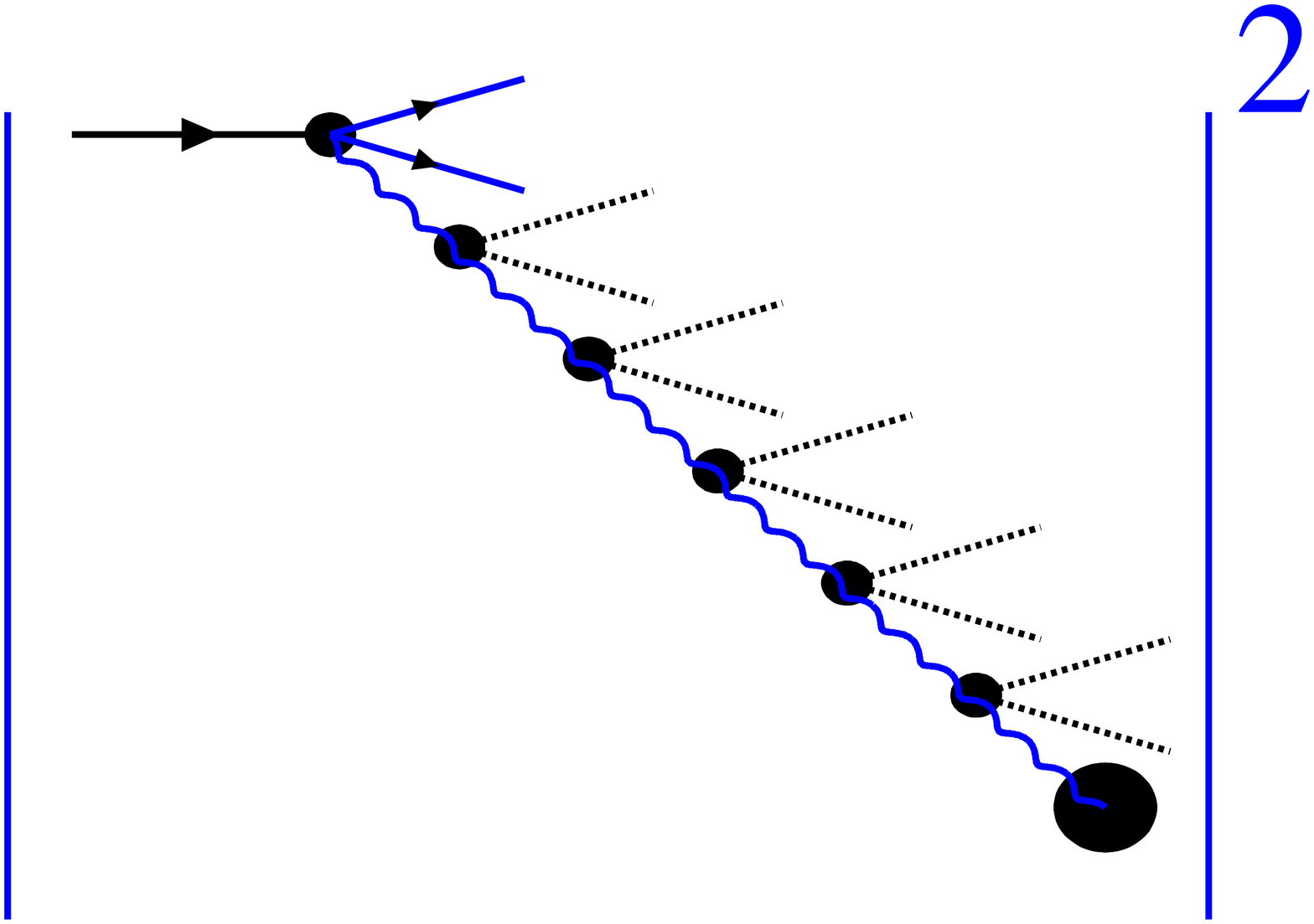,width=8cm,height=5.5cm} &
\epsfig{file=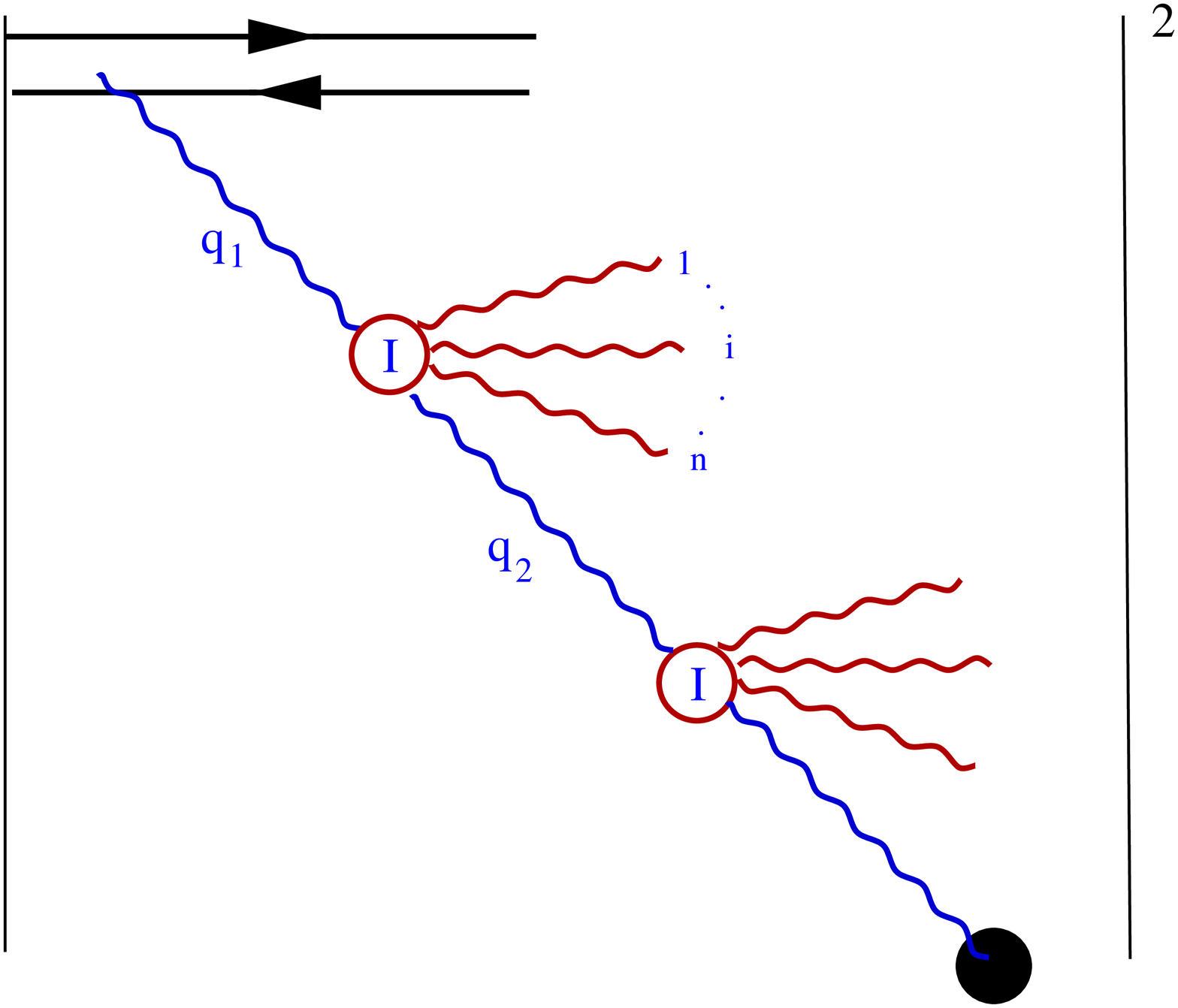,width=8cm,height=5cm}\\
Fig.2-a & Fig.2-b\\
\end{tabular}
\caption{}
\end{figure}

The physical interpretation is very simple:
 t - channel gluon scatters off { \it
semi-classical vacuum gluon fields} (see Fig.2-a). This scattering leads
to an
excitation of
the quark zero modes \,\,$\longrightarrow$\,\, pion
production. In general, This picture is very close to one that J. Bjorken
has guessed \cite{BJ}.

\section{QCD instantons and the soft
Pomeron}
The picture, that has been discussed, suggests two key observation: (i)  a
soft Pomeron is a coherent state of strong semi-classical gluon field
$F_{\mu \nu} \,\propto\, 1/g(M_0) \,\gg\,1$;  and (ii) the typical
momentum scale for these fields is rather large $M_0 \propto 1/g(M_0)$.
A natural candidate for such fields is an instanton contribution. It is
well known that instantons are the only classical solutions of QCD which
describe the transition between different QCD vacua \cite{SHU}. We found
instructive to view a Pomeron\cite{KKL,BKKL}  as a process shown in
Fig.2-b, in which
 t - channel gluon propagates through space {\it 
inducing a chain of instanton transition} between different vacua and
producing more gluons in each transition.  Practically, it means that we 
calculate a ladder digram of Fig.3-a with the vertices defined in Fig.3-b.
Our main assumption is that interactions between instantons result in
diluted  gas of instantons  with
 a definite size \cite{SHU}   $\rho
\,\,=\,\,\rho_0\,\,\approx\,\,0.3\,\,fm$. We see a strong support of this
hypothesis in lattice calculations (see Ref.\cite{RS} and references
therein).  We can reduce the problem of high energy asymptotic to the
ladder diagram of Fig.3-a using the following parameters:
\be 
\alpha_S(\rho_0 ) \,\,\ll\,\,1\,\,\,\,\, &\,\,\,\,\,  e^{-\,\frac{2
\,\pi}{\alpha_S(\rho_0)}}\,\,\ll\,\,1 \,\,;\label{P1}\\
e^{-\,\frac{2 \,\pi}{\alpha_S(\rho_0)}}\,\,\ln s \,\,\geq\,\,1\,\,\,\,\, &
\,\,\,\,\,
\frac{\rho_0\,\,M_0\,}{\alpha_S}\,\geq\,\,1\,\,.\label{P2}
\ee
The ladder diagram sums $ (  e^{-\,\frac{2 \,\pi}{\alpha_S(\rho_0)}}\,\ln
s)^n \left(
\frac{\rho_0\,
M_0 }{\alpha_S} \right)^m
 $
contributions. 

\begin{figure}[htbp]
\begin{tabular}{c c}
\epsfig{file=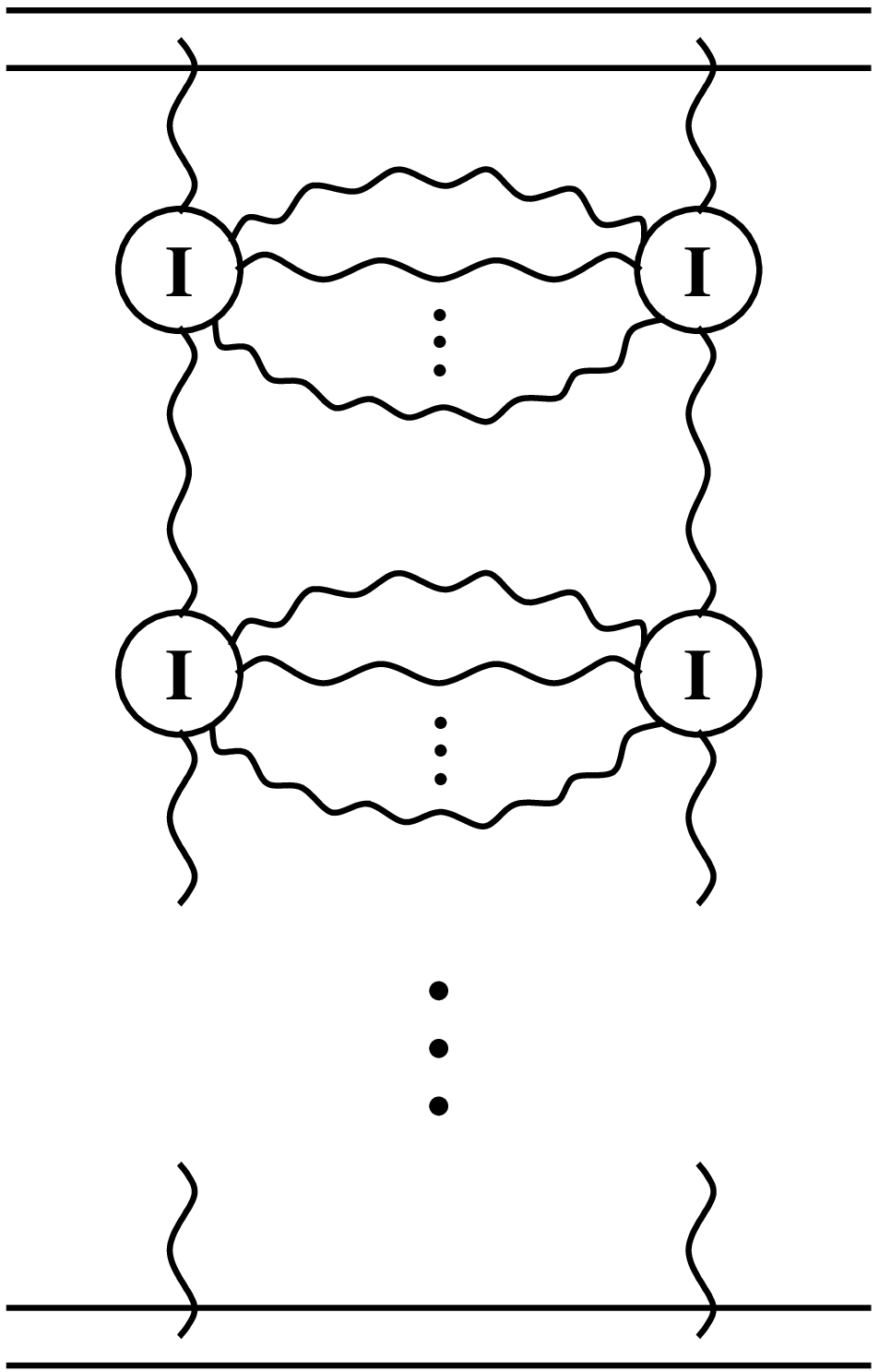,width=5cm,height=8cm} &
\epsfig{file=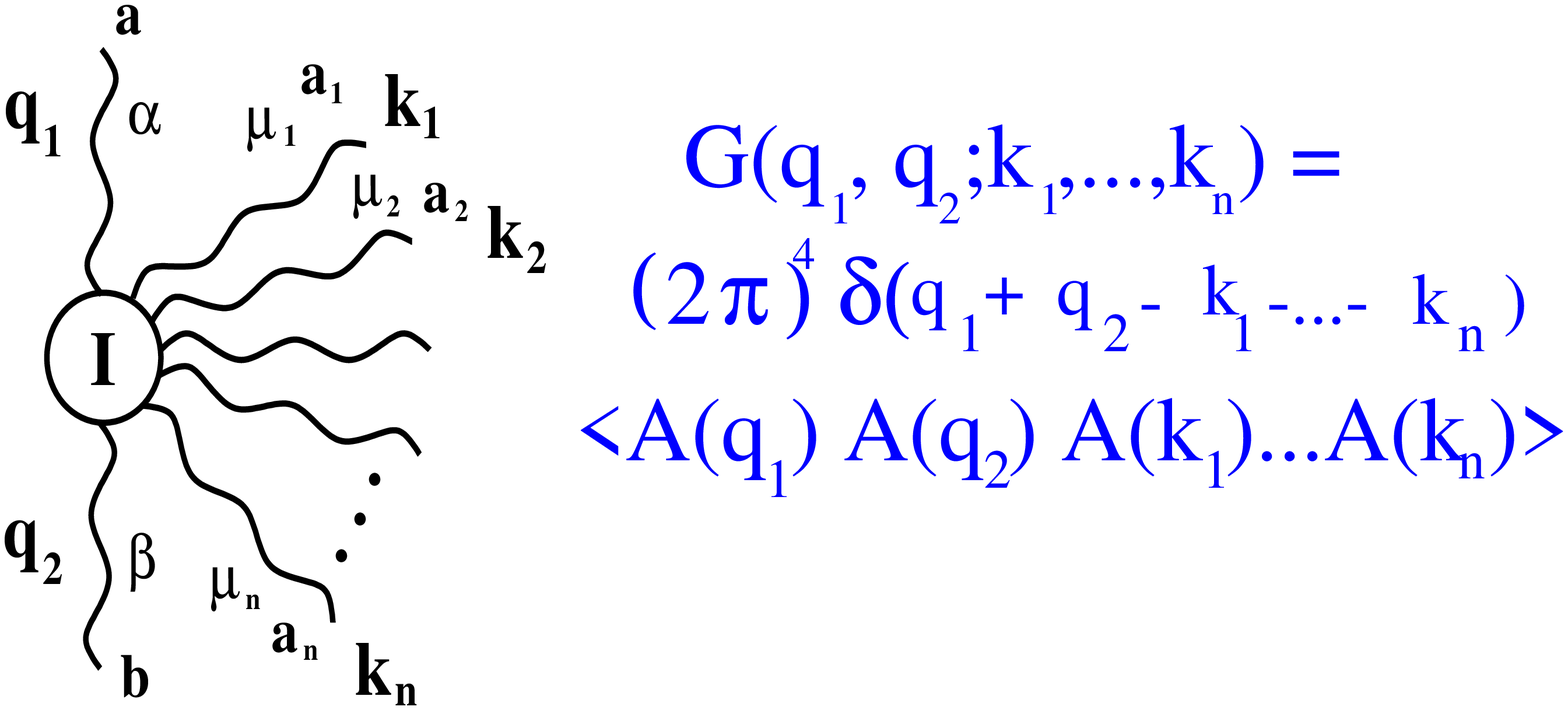,width=11cm,height=6cm}\\
Fig.3-a & Fig.3-b\\
\end{tabular}
\caption{}
\end{figure}
Our hope is that $\Delta_P \,\,\,\propto\,\,\,\rho^4_0\,\,\epsilon_{vac}
\,\,\approx\,\,0.016$ (see Fig.4).
\begin{figure}[htbp]
\begin{tabular}{l l }
\epsfig{file=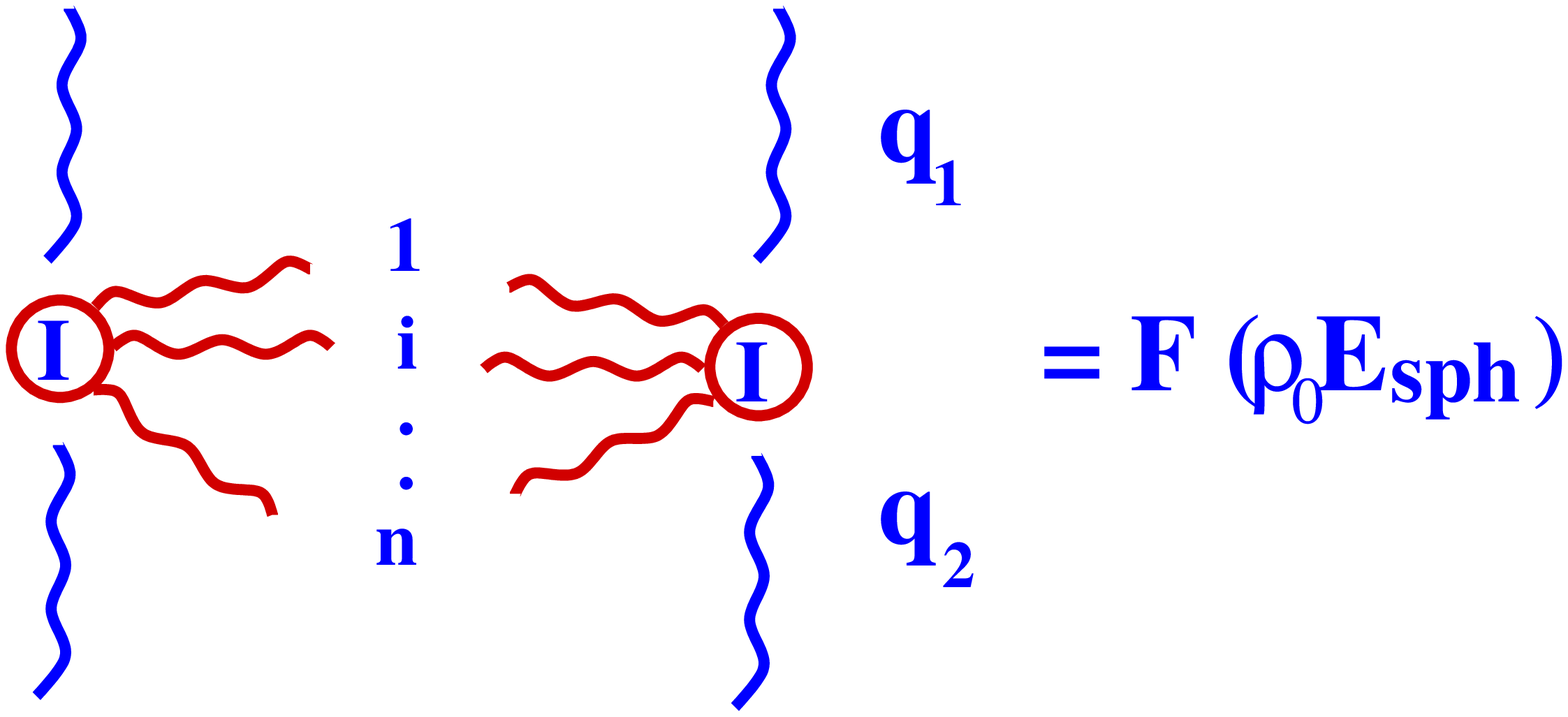,width=12cm,height=6cm} &
Fig.4-a \\
\epsfig{file=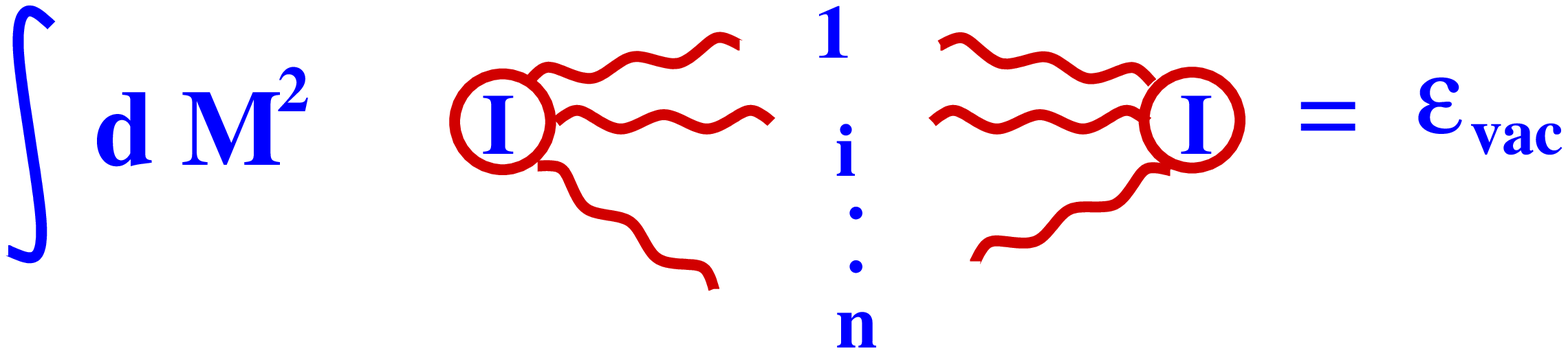,width=12cm,height=3cm} &
Fig.4-b\\
\end{tabular}
\caption{}
\end{figure}
Actually, the value of intercept $\Delta_P \propto  \int^{M^2_0}\, d M^2
\,\,\sigma( G + G\, \rightarrow\, I\,
 \rightarrow nG )$. We  believe that $M_0 \approx E_{sph}$ and
$ \sigma( G + G \, \rightarrow \,I\, \rightarrow nG )
\,\,\,
\propto \,\,\, \exp \left(\,\,\frac{1}{\alpha_S} \,F( \frac{M}{E_{sph}})
\,\right)$. The form of the ``holy grail" function $F( \frac{M}{E_{sph}})$
is  not known. In our calculation we just cut integral at $M_0 \approx
E_{sph}$. The result strongly depends on the value of cutoff. However, out
main idea is to find a relation between the value of the Pomeron
intercept and  the vacuum energy ( see Fig.4-a and Fig.4-b). We have not
found such a relation yet and we just calculate the intercept using $M_0
=2.4 GeV$. Fig.5-a gives an answer to the question why the
non-perturbative intercept is so small. One can see that the value of the
intercept has a maximum as function of $\alpha_S$. As it has been
mentioned we cannot guarantee the value of $\Delta_P$ but our approach
provides an  explanation how this value could be small even for large
$\alpha_S$. Fig.5-b shows the calculated Pomeron trajectory which turns
out to be nonlinear. It is  interesting to notice that the slope of the
Pomeron trajectory  $\alpha'_P = \Delta_P \times 2.0 GeV^{-2}$ in a good
agreement with the experimental data.

\begin{figure}[htbp]
\begin{tabular}{c c}
\epsfig{file=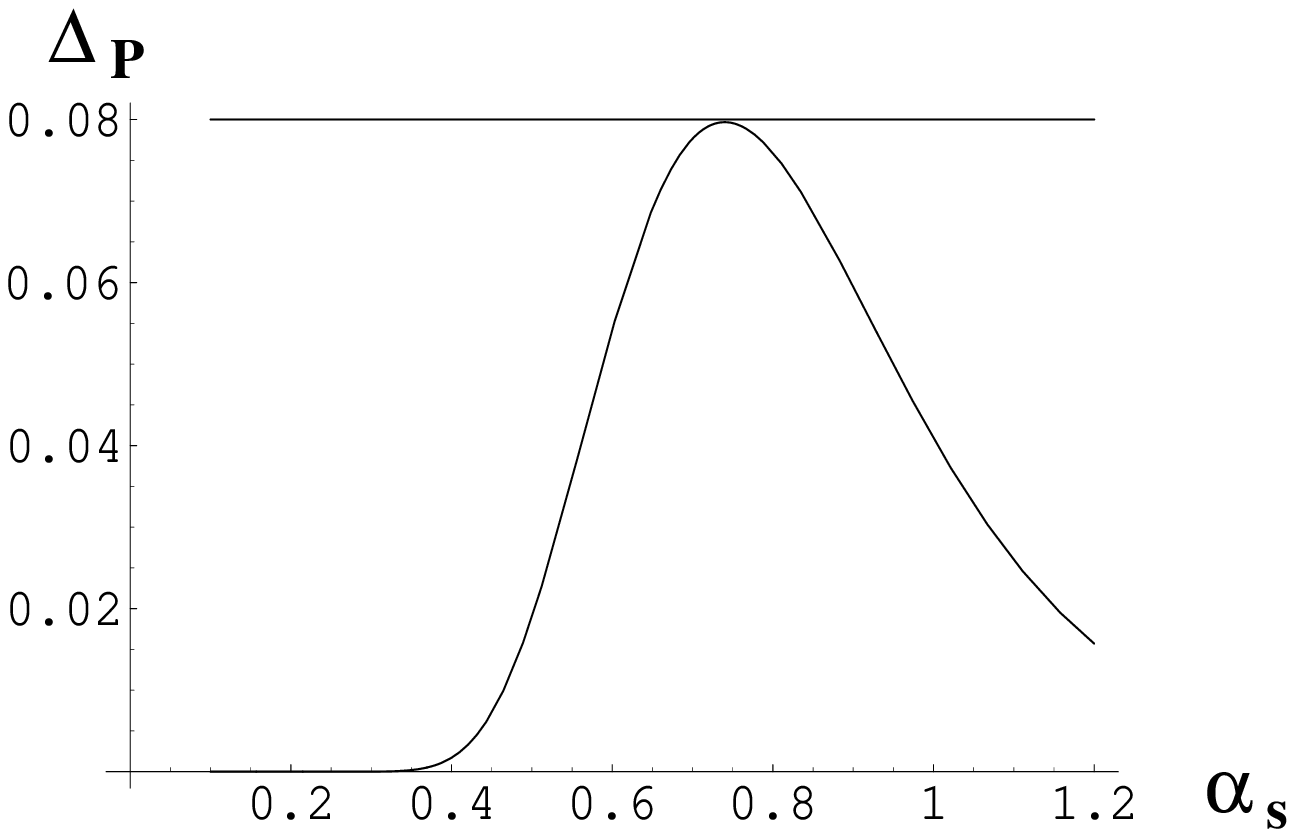,width=8cm,height=7cm} &
\epsfig{file=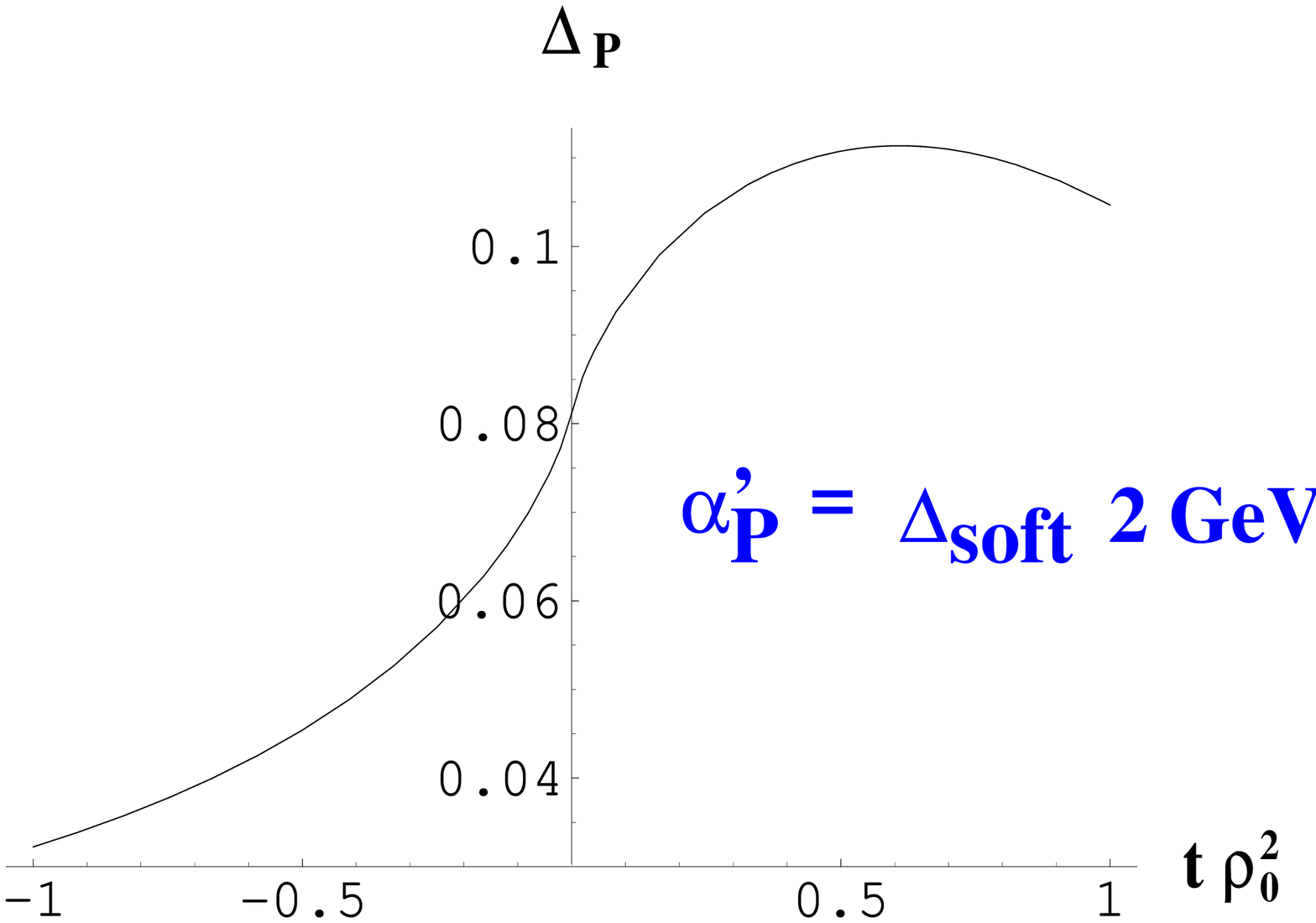,width=8cm,height=7cm}\\
Fig.5-a & Fig.5-b\\
\end{tabular}
\end{figure}

\section{Resume}
Concluding this talk I would like to repeat our point of view: the soft
Pomeron has a close relation to the structure of QCD vacuum with typical
semi-classical strong fields $ A_{\mu} \,\,\,
\propto\,\,\,1/g$ and typical large  momentum scale $M_0\,\, = \,\,E_{sph}
\,\, \propto\,\, 1/\rho_0 \alpha_S $. We demonstrate that such strong
field give rise to Reggeon with  reasonable intercept. We believe that all
other contributions are still small at scale $M_0$ since $\alpha_S(M_0)
\ll 1$. 

The difference of our approach from others on the market \cite{OA} is
that there is no   clear relation to
property of QCD vacua in them, while we have such a relation in our
approach.

We want to stress that our approach as well as other ones is based on the
assumption that long distance physics which provides confinement of quark
and gluons cannot lead to a Pomeron contribution. The reason for this
assumption is the only one: our phenomenological Pomeron has a
sufficiently large momentum scale.

We hope that our approach will trigger a return to theoretical discussion
of the Pomeron structure  with direct   addressing to an achieved
knowledge
of non-perturbative QCD. 
 
{\bf Acknowledgments:}

I wish to thank Ia. Balitsky,  E. Gotsman, D. Kharzeev,Yu. Kovchegov, U.
Maor
 and C-I Tan  for a pleasure to work with them and to discuss with them
all
problems  of ``soft" physics.

This research was supported by part by Israel Academy of Science and
Humanities and by BSF grant \#
98000276.

\end{document}